\documentclass[a4paper,11pt]{article}
\usepackage[T1]{fontenc}
\usepackage{color}
\usepackage{euscript}
\usepackage{amsmath,amssymb,amsthm}
\usepackage{amsfonts}
\usepackage{enumerate}
\usepackage{graphicx}
\usepackage{graphics}
\usepackage{hyperref}
\usepackage{natbib}

\def\1{1\!{\rm l}}
\def\argmin{\mathop{\mathrm{argmin}}}

\newcommand{\mubf}{\text{\mathversion{bold}{$\mu$}}}

\newcommand{\Xbf}{\mathbf{X}}
\newcommand{\Ybf}{\mathbf{Y}}

\newcommand{\Tbf}{\mathbf{T}}

\def\1{1\!{\rm l}}
\newtheorem{remark}{Remark}

\title{Segmentation of
multiple series using a Lasso strategy}
\author{Karine Bertin \thanks{CIMFAV-Facultad de Ingenier\'ia, Universidad de Valpara\'iso, Valpara\'iso, Chile}, Xavier Collilieux \thanks{IGN LAREG, Universit\'e Paris Diderot Sorbonne Paris Cit\'e, Paris, France ; Observatoire de Paris, SYRTE, CNRS, UPMC, Paris, France}, Emilie Lebarbier \thanks{AgroParisTech UMR518, Paris 5e and INRA UMR518, Paris 5e, FRANCE} and Cristian Meza \thanks{CIMFAV-Facultad de Ingenier\'ia, Universidad de Valpara\'iso, Valpara\'iso, Chile}}

\date{}

\begin{document}

\maketitle

\begin{abstract}
We propose a new semi-parametric approach to the joint
segmentation of multiple series corrupted by a functional part. This problem appears in particular in geodesy where GPS permanent station coordinate series are affected by undocumented artificial abrupt changes and additionally show prominent periodic variations. Detecting and estimating them are crucial, since those series are used
to determine averaged reference coordinates in geosciences and to infer small tectonic motions induced by climate change. 
We propose an iterative procedure based on Dynamic Programming for the segmentation part and Lasso estimators for the functional part. Our Lasso procedure, based on the dictionary approach, allows us to both estimate smooth functions and functions with local irregularity, which permits more flexibility than previous proposed methods.
This yields to a better estimation of the bias part and improvements in the segmentation. The performance of our method is assessed using simulated and real data. In particular, we apply our method to data from four GPS stations in Yarragadee, Australia. Our estimation procedure results to be a reliable tool to assess series in terms of change detection and periodic variations estimation giving  an interpretable estimation of the functional part of the model in terms of known functions.

\end{abstract}

\section{Introduction}

The objective of segmentation methods is to detect abrupt changes
(called breakpoints) in a signal. Such
segmentation problems arise in many areas: in biology for the
detection of chromosomal aberrations (\citealp{PRL05};
\citealp{LJK05}), in meteorology and climate (\citealp{CM2004}) to homogenize temperature and precipitation series or
in geodesy for the detection of changes in GPS location series
\citep{W2003}. In the latter example, none of the currently used segmentation methods have been shown to perform best than time series visual inspection (\citealp{Gaz}). One of the motivations of this paper is to develop an automatic method to tackle the analysis of such type of data.

GPS permanent stations that continuously monitor their coordinates have been deployed all over the world for more than 20 years. Their three-dimensional coordinate series are usually post-processed by scientists from raw code and phase observations at a daily or weekly basis, yielding series up to 1000 or 7000 records with a typical precision of a few millimeters. Such series are used to determine accurate station velocities for tectonic and Earth's mantle studies, with a typical magnitude of a few millimeters per year to about ten centimeters per year (\citealp{King2010}). Such long-term coordinates (mean positions and velocities) of a worldwide network of  stations also materialize a Terrestrial Reference Frame which is used for mapping purposes or for studying slowly varying physical phenomena including sea level variations (\citealp{ALT}). In addition, coordinate time series themselves were analyzed to infer information about ice melting and climate change (\citealp{Wu2010}; \citealp{Wu2011}; \citealp{Wahr}).

The observed coordinate variations reflect the ground deformations at the station including tectonic signals (generally a trend, mostly in the horizontal components) as well as environmental signals from the vicinity of the station, such as soil moisture or atmospheric pressure changes. The latter could be approximated by periodic signals with dominant annual and semi-annual periods (\citealp{DFCM2002}). The observational noise exhibits more autocorrelation at long periods (\citealp{WBFJNP2004}) but specific systematic errors of small magnitudes also show up at some well known periods, which are either submultiple of 350.5 days (\citealp{RACD2008}) or annual like thermal deformation of the station monumentation and the ground. Abrupt changes from a few millimeters to meters are superimposed to those variations. They are related to instrumental changes (documented or not), GPS multiple signal reflection, earthquakes or changes in the raw data processing strategy. The detection of these offsets but also of the periodic components is fundamental for the above mentioned applications. Up to now, offsets are first identified visually and the periodic components are estimated in a second run for interpretation (\citealp{INT}).

It is common to observe the same situation in genomics for the detection of chromosomal aberrations since the
biological phenomenon (corresponding to the segmentation) can be
contaminated by a probe effect or a wave-effect (see
\citealp{PLHRTR11} and references therein). Neglecting these effects
could generate false detection and leads to wrong conclusions about
the aberrations. As illustrated in Section \ref{sec:study}, other examples can be found where a set of biases represented by some functions needs to be adjusted within a segmentation model.\\

In all these data, the form of
the functional biases are not always well specified, or are even unknown. Using a non-parametric approach is very useful since it does
not require specification of the form of the functions to estimate.
In this sense, \cite{PLHRTR11} proposed a semi-parametric approach to the
joint segmentation of multiple series in the genomic application
field. When the segmentation is specific to each series and the biases (probe effect or wave-effect) are shared by
all the series, considering multiple series allows them
to better estimate these biases and so to improve the segmentation.
The model they proposed is split into two parts: a parametric
part corresponding to the segmentation and a non-parametric part (the
functional one) which is estimated using wavelets, splines or is
viewed as a fixed effect.  On the one hand, the estimation with spline or wavelet  gives good results when the biases are smooth functions but  fails when these present local irregularities.  On the other hand, the approach with the fixed
effect model tends to catch both local irregularities of the bias
and the noise, which can produce erratic estimation of the bias
part.\\

In this article, we propose a more flexible modelization of the
functional part by estimating it using a dictionary approach. In
other words, it is estimated by linear combinations of functions
with different regularities: smooth functions (for example spline
functions or Fourier functions) and more irregular functions  (for
example spiky functions). To select the relevant functions, we use a
Lasso-type strategy introduced by \cite{lasso} and recently applied
in a semi-parametric framework by \cite{ABMR2012} resulting in an
estimation procedure with good practical and theoretical performance
(oracle-type estimator). Lasso non-parametric estimators have
several advantages. A first one is that the size of the dictionary can be
large and this does no affect the computational cost of the method.
As a consequence, many different functions can be put in the
dictionary. This method is then very flexible and
allows us to estimate functions with both smooth components and local irregularities.  Moreover
the resulting estimators are sparse linear
 combinations of the functions of the dictionary. In practice this is helpful for the interpretation of the results. \\

As usual in the segmentation context with a maximum penalized
likelihood estimation framework, we
first estimate the segmentation parameters and the
non-parametric part, the number of segments being fixed. Then
we apply a model selection strategy to choose this number. For the first task, the two parts can not be estimated simultaneously. Indeed in order to infer the breakpoint parameters, it is now well known that Dynamic
Programming (DP) strategies remain among the most efficient. However this algorithm can only be applied when the contrast to be optimized is additive with respect to the segments (\citealp{BP03,CM2004,PRL05}). And this is not the case when there is a global parameter, as the function in our model \citep{BP03}. This is why, following \cite{PLHRTR11} or \cite{BP03}, our method consists in an
iterative two-steps procedure which alternates between the
segmentation issue and a Lasso-type estimation of the functional part.

We apply this strategy to simulated data where the functional part is a mixture of smooth functions and irregular functions. We obtain good results for
both segmentation and functional bias parts and, in particular, we outperform the methods of \cite{PLHRTR11} with wavelet, spline or fixed effect.
Moreover we apply our method to GPS data from Australian stations. 
For these data, we find several breakpoints of interest. The estimated
non-parametric part is found to be relevant since the obtained
periodic functions have been suggested in previous studies. Their
amplitudes and phases are more relevant for geophysical
interpretation (see for example \citealp{DFCM2002};\citealp{INT} for such an
interpretation), since they have been  simultaneously estimated all together and jointly with the segmentation part. \\

This article is organized as follows. In Section~\ref{sec:model}, we
present the semi-parametric segmentation model for multiple series.
In Section~\ref{sec:procedure}, we describe our two-step iterative
procedure based on DP for segmentation part and Lasso dictionary
approach for the functional part given a fixed number of segments,
and the model selection strategy for choosing the number of
segments. In Section~\ref{sec:study}, a simulation study
is carried out to assess the performance of our method comparatively to other methods and we illustrate the improvements obtained for a real climatic data set. In Section~\ref{sec:appli}, we apply our method to the geodetic data described above and a final conclusion is given in Section~\ref{sec:conclu}.

\section{Semi-parametric model}\label{sec:model}
We observe $M$ series. We note $y_{m}(t)$ the observed signal of the series $m$
at time $t$ and we suppose that it satisfies for $m\in\{1,\ldots,M\}$
\begin{equation} \label{ProposedModel}
y_{m}(t)=\mu_{m}(t)+f(x_{m}(t))+ e_{m}(t),
\end{equation}
where $\mu_{m}(t)=\mu^{m}_{k}$  if $t \in I_k^{m}=(\tau^{m}_{k-1},\tau^{m}_k]$,
$x_{m}$ represents possible covariates (the simple one is the time $t$), $f$ is an unknown function to be estimated, $\tau^{m}_k$ is the $k$th breakpoint of the series $m$, $\mu_{k}^{m}$ is the mean of the series $m$ on the segment $I_k^{m}$ and the $e_{m}(t)$ are i.i.d centered Gaussian with variance $\sigma^2$. We note $K_m$ the number of segments of the $m$th series and $K=\sum_{m=1}^M K_{m}$ the
total number of segments. Note that the segmentation is specific to each series. \\
For $m\in\{1,\ldots,M\}$, the series $m$ has $n_m$ observations in the times $t_{mi}$, $i\in\{1,\ldots,n_m\}$, so the total number of observations is $N=\sum_{m=1}^M n_m$ and the model is
\begin{eqnarray} \label{Observations}
y_{mi}&=&\mu_{mi}+f(x_{mi})+ e_{mi}, \\
&&\forall i\in\{1,\ldots,n_m\}, m\in\{1,\ldots,M\},\nonumber
\end{eqnarray}
where
$y_{mi}=y_m(t_{mi})$, $x_{mi}=x_m(t_{mi})$, $e_{mi}=e_m(t_{mi})$ and $\mu_{mi}=\mu_{m}(t_{mi})$. We define the vectors  $y_m:=(y_{mi})_i$, $x_m:=(x_{mi})_i$, $e_m:=(e_{mi})_i$ and $\mu_m:=(\mu_{k}^{m})_k$. \\
The parameters of the model are the means $\mu^{m}_{k}$, the
breakpoints $\tau_{k}^m$, the function $f$, the variance $\sigma^2$
and the number of segments $K$.

\section{Estimation procedure}\label{sec:procedure}

As usual in the segmentation estimation framework, the parameters are estimated for a fixed number of segments $K$ for which we propose here a DP-Lasso procedure, then $K$ is choosen using a model selection strategy.

\subsection{A DP-Lasso estimation procedure}

 Following \cite{BP03}, we propose an iterative procedure that alternates the segmentation part with the estimation of $f$. The function $f$ corresponds to a bias part common to each series and our objective is to estimate it non-parametrically using a Lasso-type method based on a dictionary approach. More specifically, we consider a collection of functions $\phi=\{\phi_1,\ldots,\phi_J\}$ and we propose to estimate $f$ by a linear combination of the functions $\phi_j$,
$$
f_{\lambda}=\sum_{j=1}^J \lambda_j \phi_j, \ \ \, \ \lambda \in \mathbb{R}^J.
$$
In order to write our estimation algorithm in a matricial form, we concatenate the means vectors $\mu_m$ in a vector $\mubf$ of size $K\times 1$. We denote by $\Tbf$ ([$N \times K$]) the incidence matrix of breakpoints $\Tbf=\text{Bloc}\left[{ \Tbf_{m}} \right]$ with $\Tbf_{m} = \text{Bloc}\left[ \1_{n_{k}^{m}} \right]$ of size ([$n_m \times K_{m}$]), and with $n_k^{m} = \tau_{k}^{m}- \tau_{k-1}^{m}$ the length of $k-$th segment for series $m$. $\Tbf \mubf$ corresponds to the segmentation part. Moreover, we concatenate the vectors $y_m$ and $x_m$ in the ([$N \times 1$]) vectors  $\Ybf$ and $\Xbf$. We denote by $F$ the [$N \times J$] matrix $F=(f_{i,j})$ where $f_{i,j}=\phi_j(\Xbf_i)$. \\

We denote by $\Tbf \mubf^{(h)}$ the segmentation estimated
parameters, $(\sigma^{(h)})^2$ the estimated variance,
$\lambda^{(h)}$ the estimated coefficients of the function $f$, and
$f^{(h)}=f_{\lambda^{(h)}}$ the estimated function $f$ at iteration
$(h)$. At iteration $(h+1)$, we get:
\begin{itemize}
\item given $\lambda^{(h)}$, the segmentation parameters $\Tbf \mubf^{(h+1)}$ are estimated by:
$$
\Tbf \mubf^{(h+1)}= \argmin_{\Tbf \mubf} \|\Ybf -\Tbf \mubf-F\lambda^{(h)}\|^2,
$$
where $\|\cdot\|$ stands for the $L_2$ norm in $\mathbb{R}^N$. The
problem is then reduced to segment $\Ybf-F\lambda^{(h)}$ into $K$
segments. In the case of joint segmentation, \cite{PLBR11} proposed
a double-stage of DP which used the multiple structure and allows us
to obtained the best segmentation of all the series into $K$
segments in a more reasonable computational time compared to the
classical DP.  \\
\item given $\Tbf \mubf^{(h+1)}$ and $\sigma^{(h)}$, the function $f$ is estimated using a Lasso-type strategy:
$$
f^{(h+1)}=f_{\lambda^{(h+1)}}
$$
where $\lambda^{(h+1)}$ minimizes
$$
\|\Ybf -\Tbf \mubf^{(h+1)}-F\lambda\|^2+2 \sum_{j=1}^J r_{N,j} | \lambda_j |,
$$
where following \cite{ABMR2012},\\ $r_{N,j}= \sigma^{(h)} \|\phi_j\|_N \sqrt{\gamma
\log{J}}$ with $\gamma>2$ and $\|\phi_j\|_N=\sqrt{\sum_{l=1}^{N}\phi_j^2(X_l)}$.\\

\item given $\Tbf \mubf^{(h+1)}$ and $f^{(h+1)}$, the variance $\sigma^2$ is estimated by
$$
(\sigma^{(h+1)})^2=\frac{1}{N}\|\Ybf -\Tbf \mubf^{(h+1)}-F\lambda^{(h+1)}\|^2.
$$
\end{itemize}
The algorithm stops when the difference between parameters of two successive iterations is smaller than $\epsilon$ ($10^{-3}$ in practice). \\
The final estimators are denoted $\hat{\tau}_{k}^m$, $\hat{\mu}_{k}^m$, $\widehat{\Tbf \mubf}$, $\hat{\sigma}^2$, $\hat{\lambda}$ and $\hat{f}=f_{\hat{\lambda}}$.

\begin{remark}
From a theoretical point of view, the condition $\gamma> 2$ ensures that the
resulting estimator of $f$ has good properties \citep[oracle performance, see][]{ABMR2012}.
However, in cases in which the Lasso estimation is performed within an iterative procedure involving the estimation of other parameters than $f$, the value of $\gamma$ may also influence the stability of the whole iterative procedure.
Then, $\gamma$ should be chosen as close as possible to $2$ while allowing for
the stability of the iterative algorithm.
\end{remark}

\subsection{Model selection}
The last issue is the choice of the number of segments $K$. We
propose here to use the modified BIC criterion proposed by \cite{ZhS07} and successfully
adapted  to the joint segmentation by \cite{PLBR11}:
\begin{eqnarray*}
&&mBIC_{\text{JointSeg}}(K)= \log{\left [\Gamma \left
(\frac{N-K+1}{2} \right)\right ]} \\
&& -\left ( \frac{N-K+1}{2} \right )
\log{SS_{\text{wg}}} +\left [ \frac{1}{2}-(K-M) \right ] \log{(N)}  \\
&& -\frac{1}{2} \sum_{m=1}^M \sum_{k=1}^{k_m} \log
{(\hat{\tau}_{k}^m-\hat{\tau}_{k-1}^m)}  ,
\end{eqnarray*}
where $SS_{\text{wg}}=\|\Ybf-\widehat{\Tbf
\mubf}-F\hat{\lambda}\|^2$.

\section{Study of the performance of the method}\label{sec:study}

In order to assess the performance of our procedure, so-called here {\it Lasso}, in Section~\ref{sec:simu}, we conduct the simulation study described below. We also propose to compare our method to the work of \cite{PLHRTR11}, where either the function $f$ is estimated using splines or $f$ is viewed as a fixed effect depending on the time $t$, i.e. $f(t)=\beta_t$. We call these two approaches {\it Spline} and {\it Position} respectively and we perform them on the simulated data using the \verb+cghseg+ R package, in particular using the \verb+multiseg+ R function. For our procedure, we develop our own functions in R using the \verb+lars+ R package to perform the Lasso estimation of $f$.
In addition in Section~\ref{illustration}, we illustrate on a climatic data set the need to model correctly
the function $f$ in order to avoid false detection in the
segmentation.

\subsection{Simulation study}\label{sec:simu}

\begin{figure}
\centering
\makebox{
\includegraphics[scale=0.5]{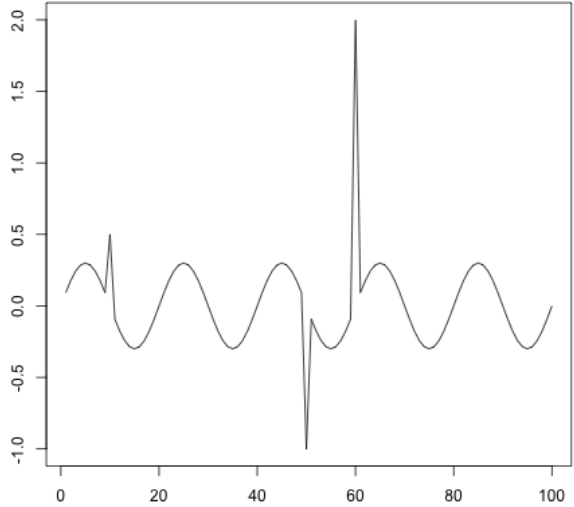}}
\caption{Simulated function $f$.}
\label{graf0.simul}
\end{figure}

\paragraph{Simulation design.} We consider the model (\ref{ProposedModel}) for series $m\in\{1,\ldots,M\}$ at time $t$:
\begin{equation}\label{ec1.simul}
y_m(t)=\mu_m(t)+f(t)+e_m(t), \quad t=1,\dots,n
\end{equation}
where $e_m(t) \sim \mathcal{N}(0,\sigma^2)$ i.i.d. The length $n$ of
the series is fixed and equal to $100$. We consider two different
numbers of series: $M\in \{10,50\}$, and five values for error
variance: $\sigma^2\in\{0.1,0.2,0.5,1.0,1.5\}$. For each series, the
number of segments $K$ follows a Poisson distribution with mean
$\bar{K}=3$ and their positions are uniformly distributed. The mean
value within each segment alternates between $0$ and a value in
$\{-2,-1,+1,+2\}$ with probability $\{0.2,0.3,0.3,0.2\}$
respectively. The function $f$ is generated as a mixture of a sine
function with three peaks (see Figure \ref{graf0.simul}):
\begin{eqnarray}\label{ec2.simul}
f(t)&=&0.3\times\sin \left(2\pi \frac{t}{20}\right)+0.5\leavevmode\hbox{1\!\rm I}_{t=0.1\times n}\\
&& -\leavevmode\hbox{1\!\rm I}_{t=0.5\times n} +2\leavevmode\hbox{1\!\rm I}_{t=0.6\times n} .\nonumber
\end{eqnarray}

Each configuration, i.e. specific values of $M$ and $\sigma^2$, is simulated 100 times.\\

For the Lasso strategy, we use a dictionary with $150$ functions: $128$ Haar functions ($t\mapsto 2^{7/2}\leavevmode\hbox{1\!\rm I}_{[0,1]}\left(\frac{2^{7}t}{100}-k\right), k=0,\ldots,2^7-1)$, the Fourier functions $(t\mapsto\sin\left(\frac{2\pi j t}{100}\right),t\mapsto\cos\left(\frac{2\pi j t}{100}\right), j=1,\ldots,10)$ and the functions $t\mapsto t$ and $t\mapsto t^2$. The Lasso estimator is obtained by LARS algorithm with $\gamma=2.1$.

\paragraph{Quality criteria.}
To study the quality of the estimation, for each configuration, we consider several criteria:
\begin{itemize}
\item
For the segmentation parameters, in order to study the global quality of the estimation, we consider the root-mean-square distance between the true mean and its estimate:\\
 $\text{RMSE}(\mu) =\left[\frac{1}{N}\sum_{m=1}^M\sum_{t=1}^{n}\left\{\mu_m(t)-\hat{\mu}_m(t)\right\}^2\right]^{1/2}$ where $N=M\times n$. Moreover, to study the performance of the estimation of the breakpoint positioning, we consider both the proportion of erroneously detected breakpoints among detected breakpoints (false discovery rate, FDR) and the proportion
of undetected true breakpoints among true breakpoints (false negative rate, FNR). \\
\item
For the function $f$, the root-mean-square distance between $f$ and its estimate:\\ $\text{RMSE}(f) =\left[\frac{1}{n}\sum_{t=1}^{n}\left\{f(t)-\hat{f}(t)\right\}^2\right]^{1/2}$ is also considered.
\end{itemize}
For each configuration, we consider the average of these criteria over the $100$ simulations.

\paragraph{Comparison between \textit{Lasso}, \textit{Spline} and \textit{Position}.} Only the results with $M=10$ are presented since the results for $M=50$ leads to same conclusions.  \\

\begin{figure}
\centering
\makebox{\includegraphics[scale=0.5]{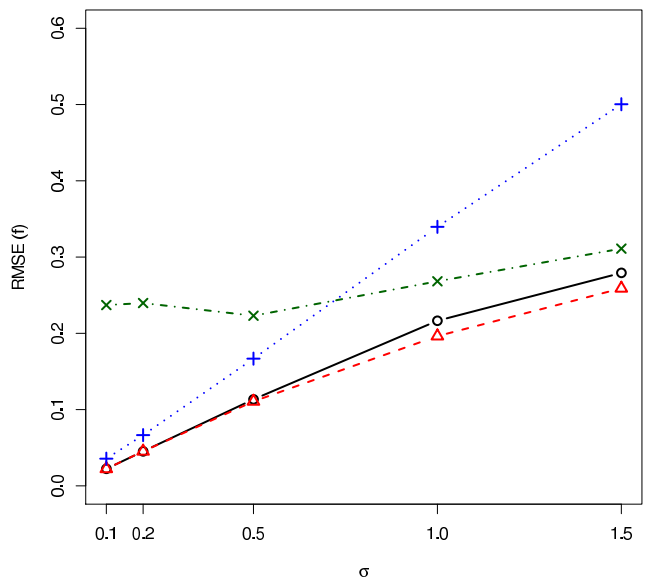}}
\caption{RMSE of $f$ with respect to $\sigma$ for \textit{Lasso} $\triangle$, \textit{Position} $+$, \textit{Spline} $\times$ and \textit{Lasso} with the true number of segments $\circ$ for $M=10$.}
\label{graf2.simul}
\end{figure}

\begin{figure}
\centering
\begin{tabular}{cc}
\makebox{\includegraphics[scale=0.4]{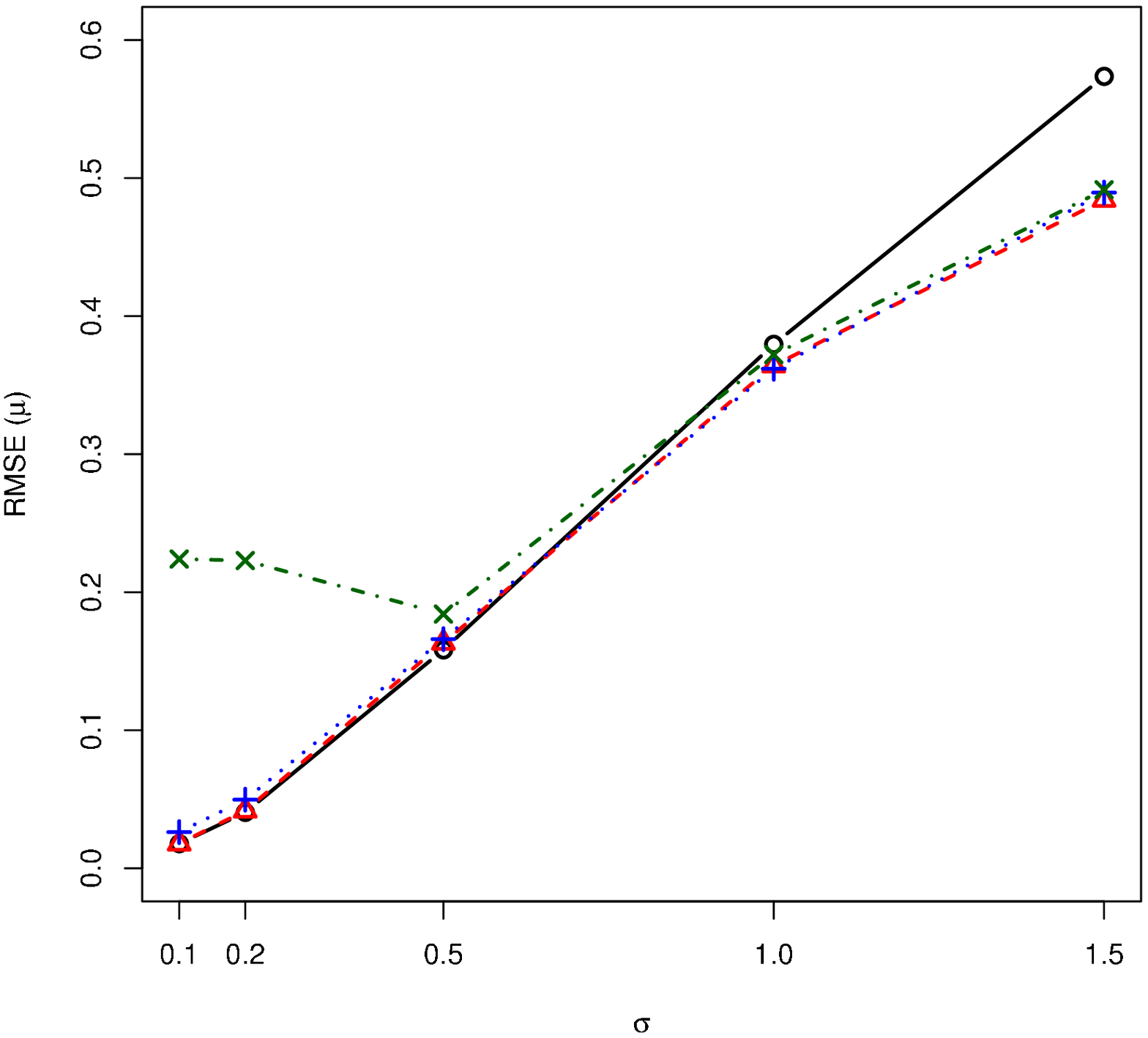}}  &
\makebox{\includegraphics[scale=0.4]{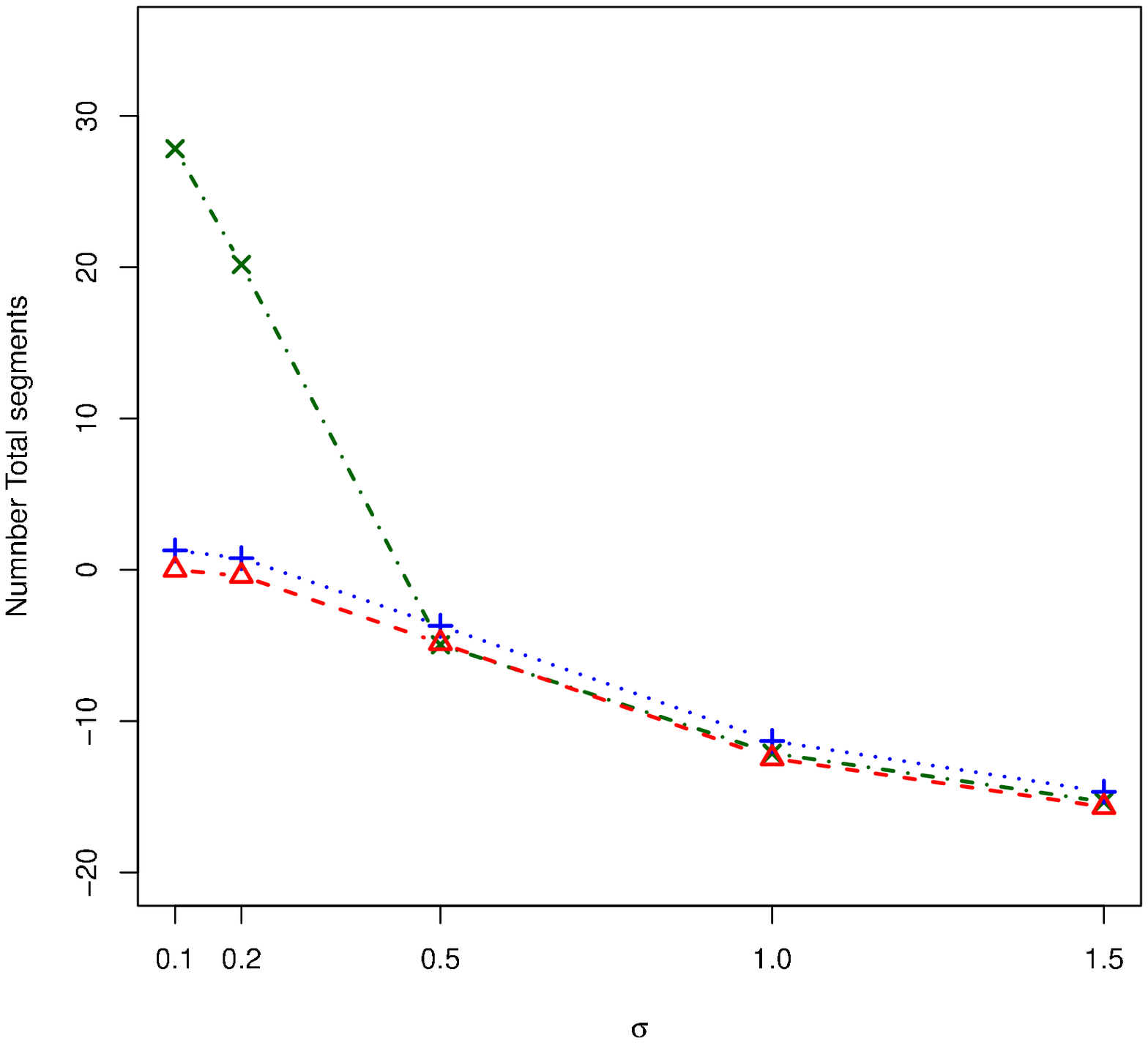}}  \\
\makebox{\includegraphics[scale=0.4]{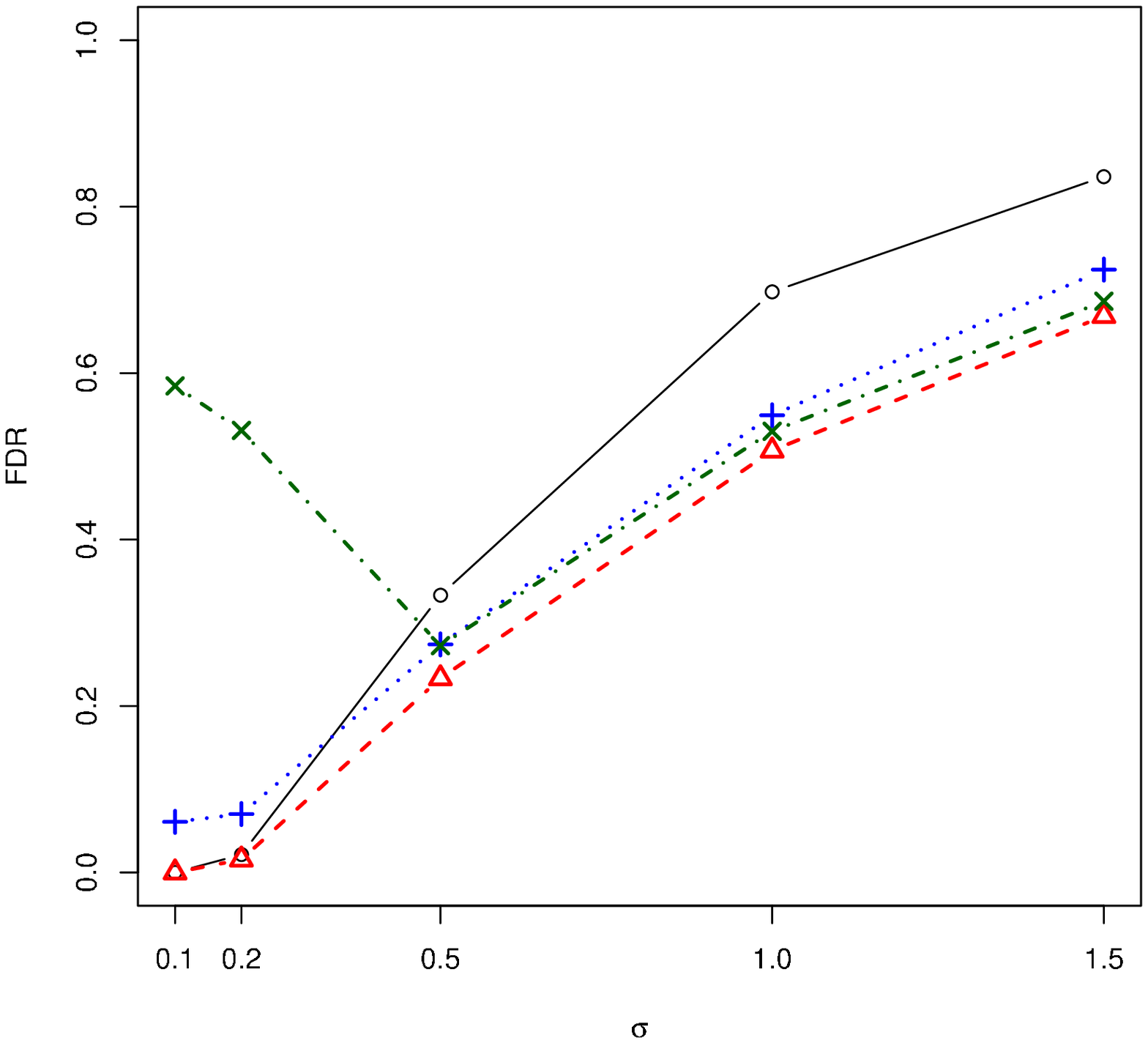} } &
\makebox{\includegraphics[scale=0.4]{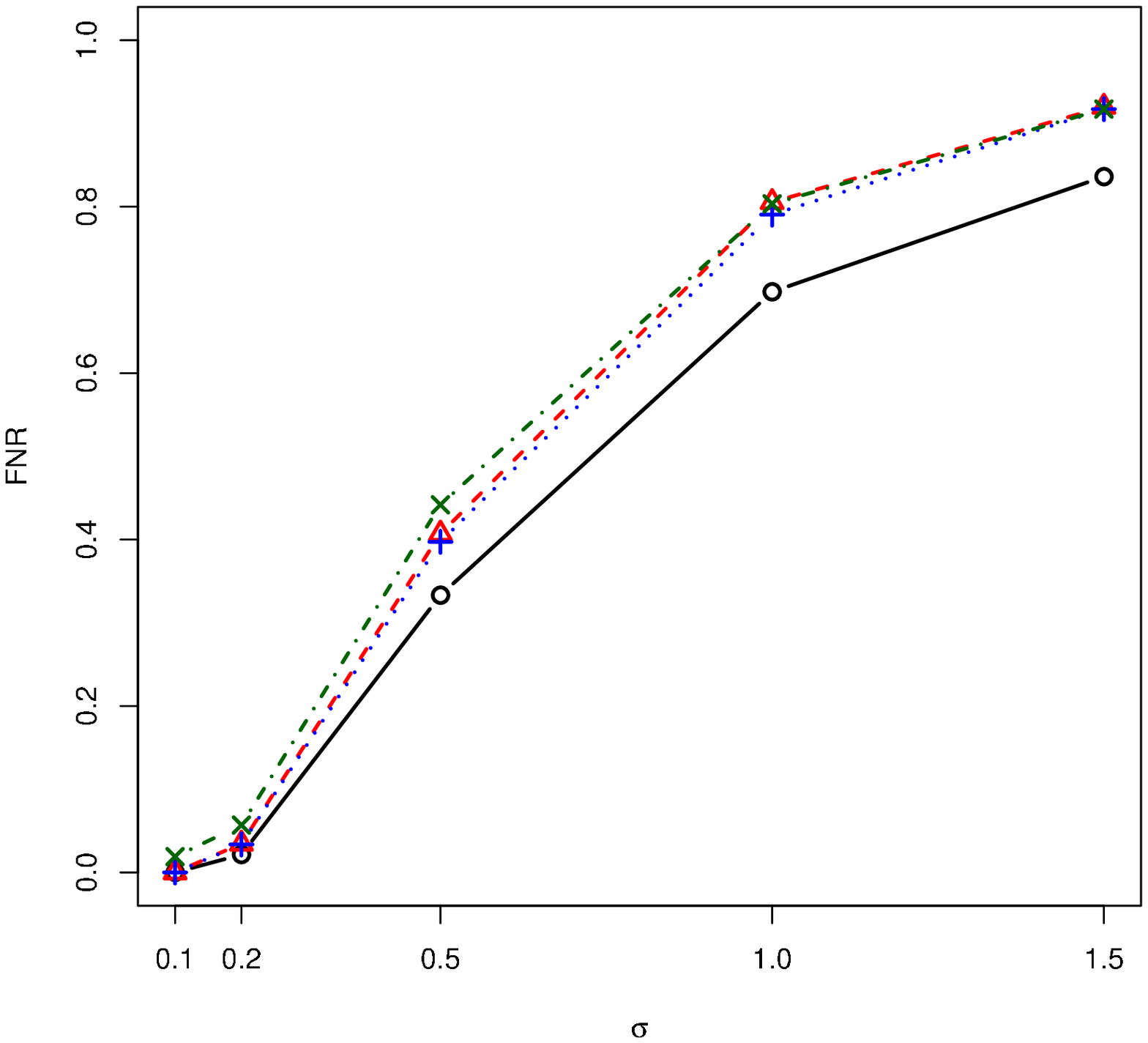}}
\end{tabular}

\caption{Results for $M=10$ with respect to $\sigma$. Top: RMSE of
$\mu$ on the left, $\hat{K}-K$ on the right. Bottom: FDR on the left
and FNR on the right. \textit{Lasso} $\triangle$, \textit{Position}
$+$, \textit{Spline} $\times$ and \textit{Lasso} with the true
number of segments $\circ$.} \label{graf3.simul}
\end{figure}

Figure \ref{graf2.simul} presents the $\text{RMSE}(f)$ for the different methods with respect to $\sigma$. We observe that the larger is the noise, the worst is the estimation of $f$ due to the confusion between the signal and the noise. Whatever the level of noise, \textit{Lasso} outperforms \textit{Position} and \textit{Spline} in terms of the non-parametric part estimation. However, the behavior of \textit{Position} and \textit{Spline} is opposite with  respect to $\sigma$. For small $\sigma$, \textit{Spline} leads to bad performances compared to \textit{Lasso} and \textit{Position}. Indeed, as expected, \textit{Spline} tends to capture the smooth part of the signal, i.e. the sinusoidal trend only, whereas the two others catch both the peaks and the trend. However, for large $\sigma$, it is more difficult to detect the peaks of the true function, resulting in closest results for \textit{Lasso} and \textit{Spline}. \textit{Position} behaves worstly since, as mentioned in \cite{PLHRTR11}, it tends to catch the trend but also the noise resulting in an erractic estimation of $f$. The bad estimation of $f$ can have consequences on the segmentation estimation. Figure \ref{graf3.simul} summarizes the results for the segmentation estimation obtained with the different methods with respect to $\sigma$. In general, \textit{Lasso} is sligthly better than the two other methods. For $\sigma$ larger than $0.5$, the results are similar, even for \textit{Position} for which  $f$ is not well estimated. However for small values of $\sigma$, since \textit{Spline} does not detect the peaks, they are considered as breakpoints in the segmentation, leading to bad results: more segments are then detected (see $\hat{K}-K$), these false breakpoints then increase the FDR and so the $\text{RMSE}(\mu)$. \\

As a conclusion, \textit{Position} and \textit{Lasso} behave similarly for the estimation of the segmentation part. The main difference concerns the estimation of $f$ which is less reliable for \textit{Position}. An important advantage of \textit{Lasso} is its flexibility in the sense that functions of different regularities can be included in the dictionary and in particular some functions chosen according to the knowlegde of the expert. The final form of the estimator $\hat{f}$ is a sparse linear combination of the dictionary functions that allows a possible interpretation of $f$ compared to \textit{Position} (see Section \ref{sec:appli}).

\paragraph{Discussion on the quality of the estimation with \textit{Lasso}.}

We first compare the results obtained with the true and estimated number of segments. In Figure \ref{graf2.simul}, we observe that the more difficult is the detection (more $\sigma$ increases), more the number of segments is under-estimated. This result was expected and is now classical in the study of model selection for segmentation. Indeed, the number of segments is reduced in order to avoid false detection. This is illustrated by a less increase of the FDR obtained with the estimated number of segments compared to the true one (Figure \ref{graf3.simul}). That leads to a better estimation in terms of segmentation (small $\text{RMSE}(\mu)$) and consequently to a better estimation of the function $f$ (small $\text{RMSE}(f)$).  \\

\paragraph{Segmentation and the estimation of $f$ as a function of the number of series.} Table \ref{table.CompM} summarizes the relative differences for two criteria, the FDR and the root-mean-square of $f$ between $M=10$ and $M=50$, for several values of $\sigma$ as:
\begin{eqnarray*}
FDR^{\sigma}&=&\frac{FDR_{10}^{\sigma}-FDR_{50}^{\sigma}}{FDR_{10}^{\sigma}},\\
RMSE(f)^{\sigma}&=& \frac{RMSE(f)_{10}^{\sigma}-RMSE(f)_{50}^{\sigma}}{RMSE(f)_{10}^{\sigma}},
\end{eqnarray*}
where, for example,  $FDR_{10}^{\sigma}$ and $RMSE_{10}(f)^{\sigma}$ denote respectively the FDR and the root-mean-square of $f$ for $M=10$ series for a specific value of $\sigma$. Table~\ref{table1.simul} shows the percentage of the true functions of the simulated function $f$ selected in the estimator $\hat{f}$ against different values of $\sigma$, with $M=10$ and $M=50$ series. The ID function corresponds to the position of the true functions in the dictionary with size $150$. Specifically, the first three functions (labels $13$, $64$ and $77$) are Haar functions centered in $10$, $50$ and $60$ and the function $137$ is the function $x\mapsto \sin \left(2\pi \frac{5 t}{100}\right)$. In addition, a FDR criterion is calculated, corresponding to the number of false selected functions among the selected ones. As expected, the increase of the number of series improves the estimation of $f$ (large $RMSE(f)^{\sigma}$). For small values of $\sigma$, the \textit{Lasso
 } procedure leads to a good performance in terms of selected functions whatever the number of series: the number of selected functions is close to the true one, and among them all the true functions of the simulated function are retrieved with less false selection (small FDR function). That leads logically to an accurate estimation of $f$ (small $RMSE(f)$ Figure \ref{graf2.simul}). For noisy configurations (large $\sigma$), fewer functions are selected, which was expected. Indeed, in this case, there are more confusion between noise and signal, the small peaks (in particular ID 13 and 64) are more difficult to detect. This is particularly true for a small number of series. Remark that for $M=50$, the ID 77 and 137 are always selected. \\
Moreover, the better accuracy of the estimation of $f$ observed for $M=50$ leads to a better positioning of the breakpoints (see $FDR^{\sigma}$). This is less marked when $\sigma$ is large.

\begin{table}
\caption{\label{table.CompM}Comparison between $M=10$ and $M=50$ series for $FDR$ and $RMSE(f)$ criteria. }
\fbox{
\begin{tabular}{*{3}{c}}
&\multicolumn{2}{c}{Relative differences} \\
$\sigma$ & $FDR^{\sigma}$ & $RMSE(f)^{\sigma}$ \\\hline
0.1 & - & 57.46\\
0.2 & 42.15 & 57.97 \\
0.5 & 9.40 & 55.58 \\
1.0 & 7.00 & 50.47\\
1.5 & 5.64 & 47.47\\
\end{tabular}}

\end{table}

\begin{table}

\caption{\label{table1.simul}Percentage, FDR and mean of number of  functions selected by Lasso.}
\fbox{%
\begin{tabular}{*{8}{c}}
\em &\em & \multicolumn{4}{c}{\em ID function}&\em FDR & \em Mean \\
&$\sigma$ & 13 & 64 & 77 & 137 &  function &length\\\hline
&0.1 & 100 & 100 & 100 & 100 & 0.052 &4.27 \\
&0.2 & 100 & 100 & 100 & 100 &  0.055 &4.29 \\
M=10 &0.5 & 26 & 99 & 100 & 100 &  0.064 &3.53 \\
&1.0 & 5 & 28 & 99 & 99 &  0.114 &2.13 \\
&1.5 & 0 & 12 & 73 & 76 &  0.137 &1.9\\\hline
&0.1 & 100 & 100 & 100 & 100 &  0.059 & 4.31 \\
&0.2 & 100 & 100 & 100 & 100 & 0.059 & 4.31\\
M=50 &0.5 & 100 & 100 & 100 & 100 &  0.068 & 4.36\\
&1.0 & 53 & 100 & 100 & 100 &  0.084 & 3.95 \\
&1.5 & 18 & 92 & 100 & 100 &  0.108 & 3.6\\\hline
\end{tabular}}

\end{table}

\subsection{Illustration} \label{illustration}

\begin{figure}
\centering
\makebox{
\includegraphics[scale=0.45]{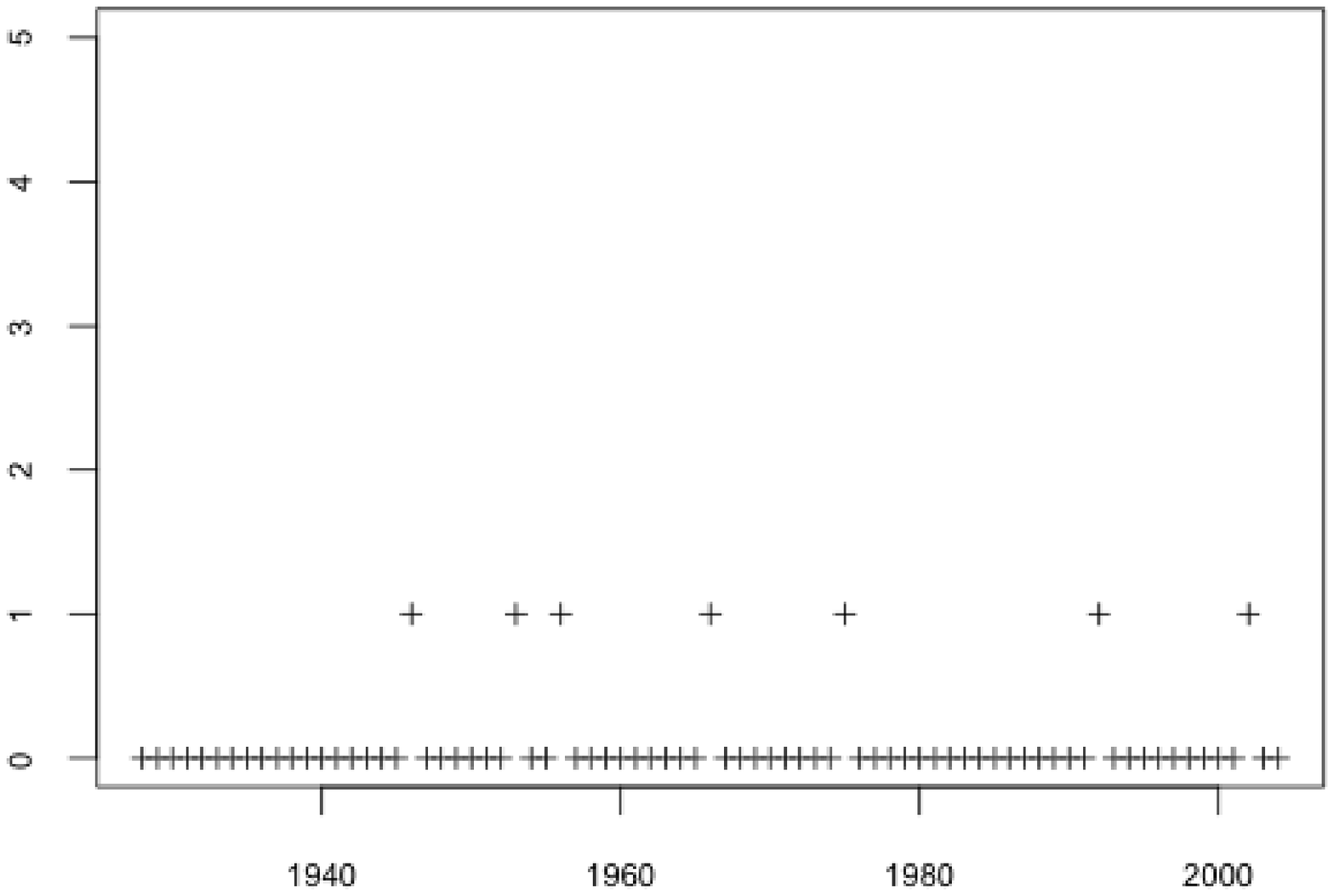}}\\
\makebox{\includegraphics[scale=0.45]{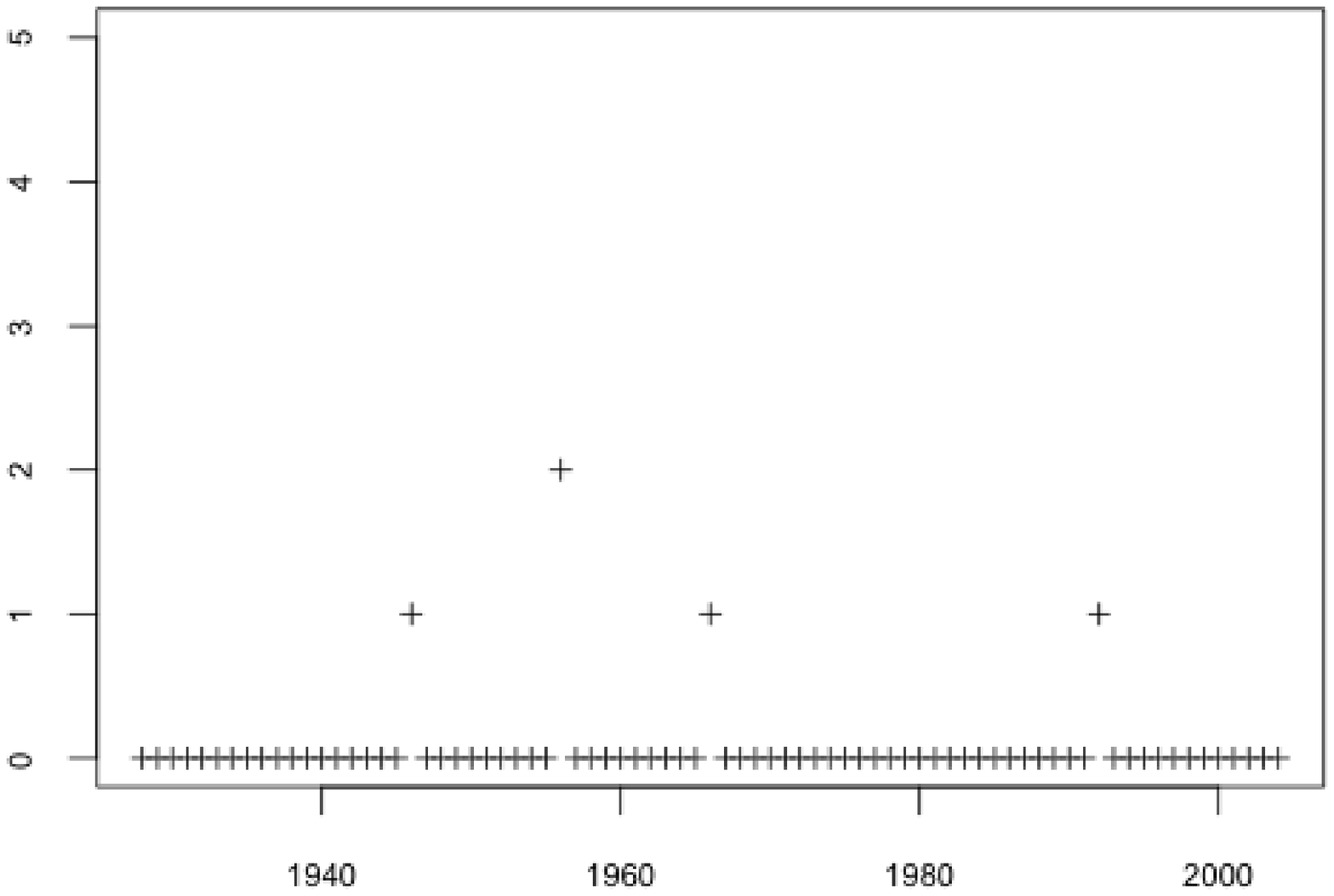}}
\caption{Number of the detected breakpoints over all the stations
obtained in case $(1)$ on the top and case $(2)$ on the bottom.}
\label{Fig1}
\end{figure}

\begin{figure}
\centering

\makebox{\includegraphics[scale=0.45]{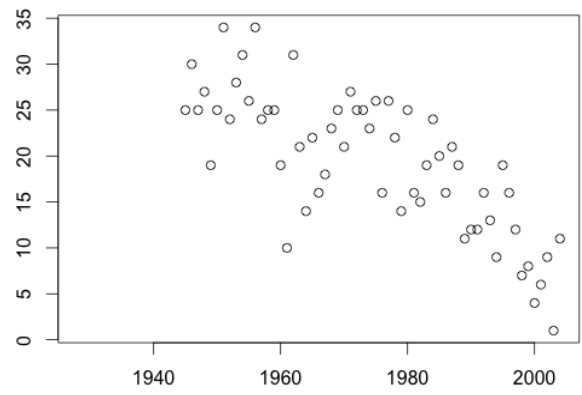}}\\
\makebox{\includegraphics[scale=0.45]{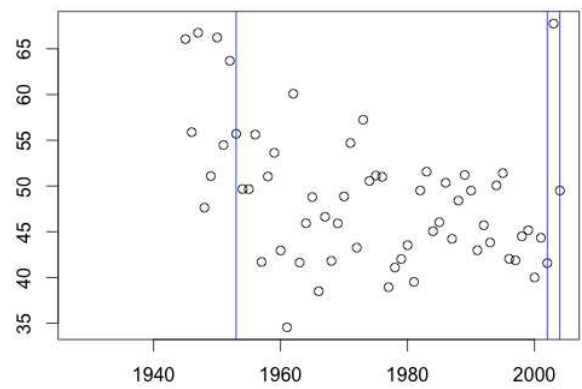}}

\caption{Top: harvest dates of the series 6.
Bottom: obtained segmentation of the corrected series in case $(1)$
(on $y_{mt}-\hat{b}x_{mt}-\hat{c}x_{mt}^2$).} \label{Fig2}
\end{figure}

\begin{figure}
\centering
\makebox{
\includegraphics[scale=0.5]{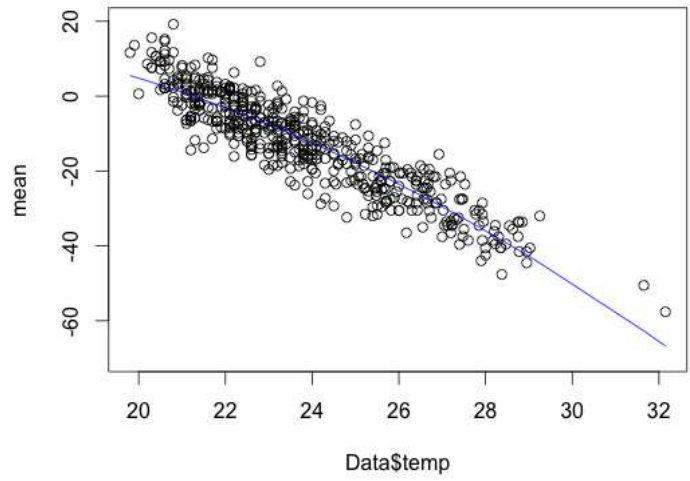}}  \\
\makebox{
\includegraphics[scale=0.5]{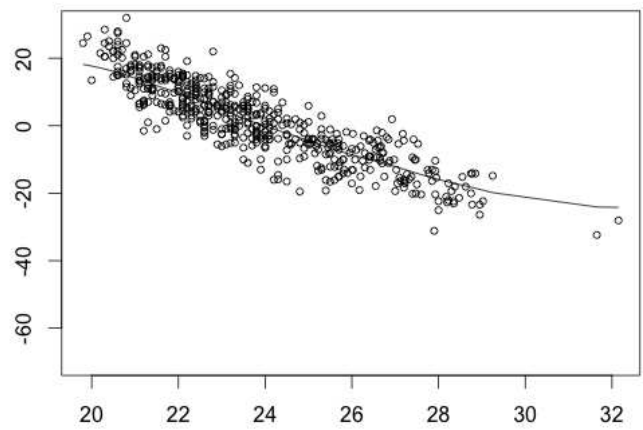}}
\caption{Fit of $f$ in case $(1)$ on the top and case $(2)$ on the bottom.}
\label{Fig3}
\end{figure}

In this section, we want to illustrate the need to model correctly
the function $f$ in order to avoid false detection in the
segmentation. To this end, we compare our procedure to the results
obtained by \cite{PLBR11} in their study on harvest dates. In this
application, the purpose is to detect changes in the agricultural
practices by detecting changes in the grape harvest dates which are
not due to the climatic effect. The data are harvest dates obtained
at 10 French stations. The model they proposed is a mixed linear
model containing a segmentation part, a random effect and a climatic
effect modelled by a degree 2 polynomial according to the
temperature. To compare with our proposed strategy, we avoid the
random effect.  The model is then written as follows:
\begin{equation*}
y_{m}(t) = \mu_{k}^m + f(x_m(t)) +e_{m}(t), \text{ if $t \in I_k^m$},
\end{equation*}
where $y_m(t)$ is the grape harvest date and $x_m(t)$ is the mean
temperature of the year $t$ for series $m$. In case $(1)$, the form
of the climatic effect $f$ is fixed to be $f(x_m(t))=b x_{mt} + c
x^2_{mt}$. In case $(2)$, no assumptions are made on the function
$f$ and it is estimated using our proposed procedure, for which we
consider a dictionary with 36 functions compound with high
resolution level Haar wavelets, Fourier basis, $x$, $x^2$ and $x^3$.
In the resulting estimator of $f$ obtained with $\gamma=2.1$, five
functions are selected. Figure \ref{Fig1} represents the number of
detected breakpoints per year over all the series for the two
models. The result obtained in case $(1)$ is slightly different from
the one obtained in \cite{PLBR11}. However, the most important
difference compared to the result obtained by our proposed procedure
concerns the year 2003 which corresponds to a very hot summer: that
year is considered as a breakpoint in case $(1)$ and not in case
$(2)$. This breakpoint appears in the series 6. Figure \ref{Fig2}
represents respectively the harvest dates of the series 6 and its
segmentation after correction in case $(1)$ (segmentation of
$y_t-\hat{b}x_{t}-\hat{c}x_{t}^2$). The temperature at year 2003
is $32.15$. As shown in Figure \ref{Fig3}, the correction of the
harvest date at this year by $\hat{b} x_{mt}+\hat{c} x_{mt}^2$ is
too strong compared to $\hat{f}(x_{mt})$ obtained in case $(2)$ that is why a false breakpoint is added (see Figure \ref{Fig2} bottom). \\

\section{Application}\label{sec:appli}

In this Section, we summarize the results obtained with our estimation procedure for the GPS dataset described in Introduction. In particular, we use the height coordinate series of four GPS stations in Australia located in Yarragadee (YAR1, YAR2, YAR3 and YARR). Those were computed by the Jet Propulsion Laboratory (JPL). They can be downloaded at\\ \footnotesize{\verb+ftp://sideshow.jpl.nasa.gov/pub/JPL_GPS_Timeseries/repro2011b/post/point/+.} \normalsize We use the series from their first observations to the 22nd of June 2013 - series provided online are updated everyday. Then the model (\ref{ProposedModel}) is considered with $M=4$ and $n_1=2862$, $n_2=5209$, $n_3=1443$ and $n_4=2197$, the respective lengths of the series.
Here they have been averaged at weekly scale. For all these series, the ground motion is assumed to be identically observed and is described with function $f(t)$. Thus, equipment changes or malfunction at individual station should show up in the segmentation. For those series, JPL detected changes using a procedure based on sequential F-test applied to the tridimensional coordinate series (M. Heflin, personal communication, 2014). \\

\begin{figure}
\centering
\begin{tabular}{cc}
\makebox{\includegraphics[scale=0.35]{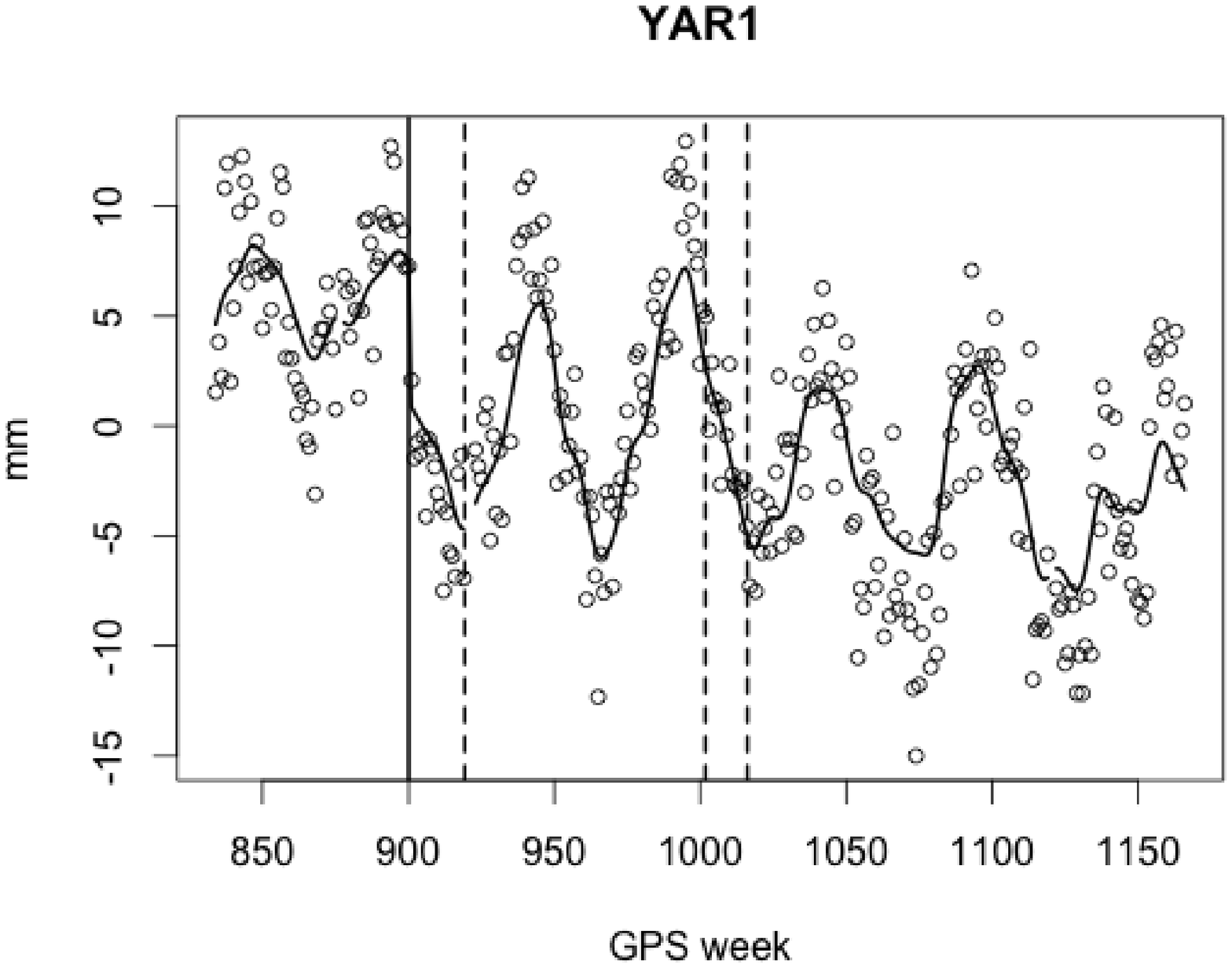}}&
\makebox{\includegraphics[scale=0.35]{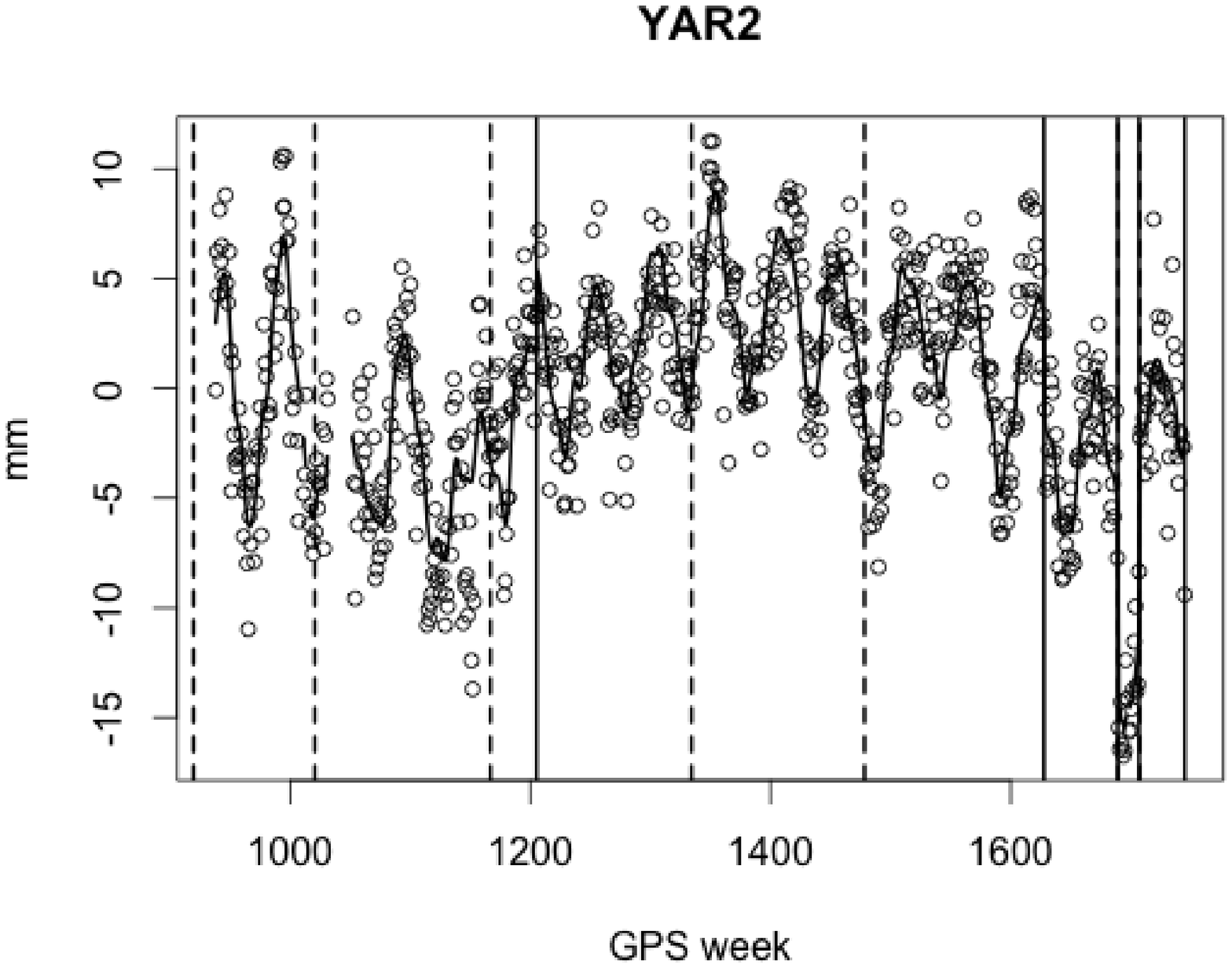}}\\
\makebox{\includegraphics[scale=0.35]{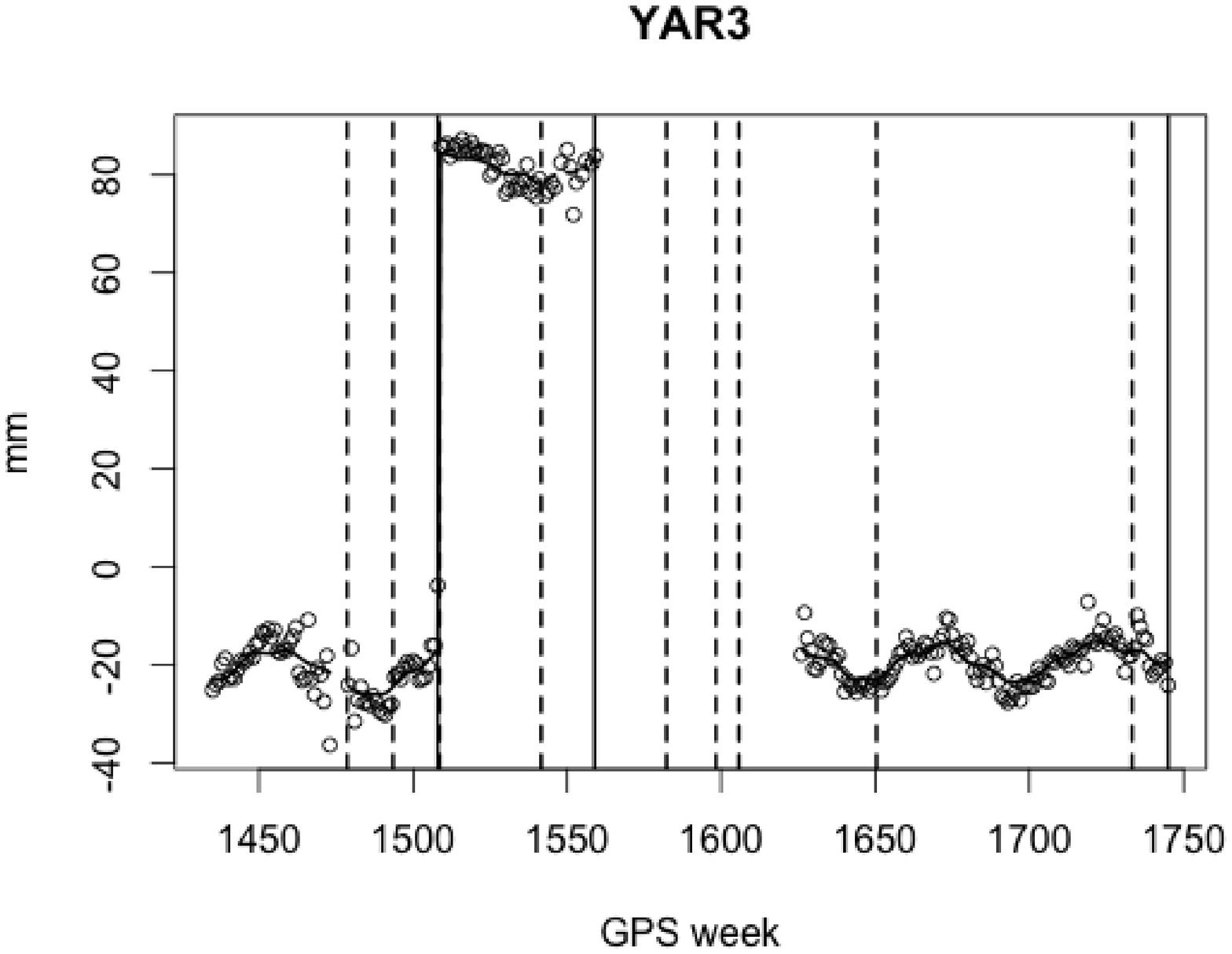}}&
\makebox{\includegraphics[scale=0.35]{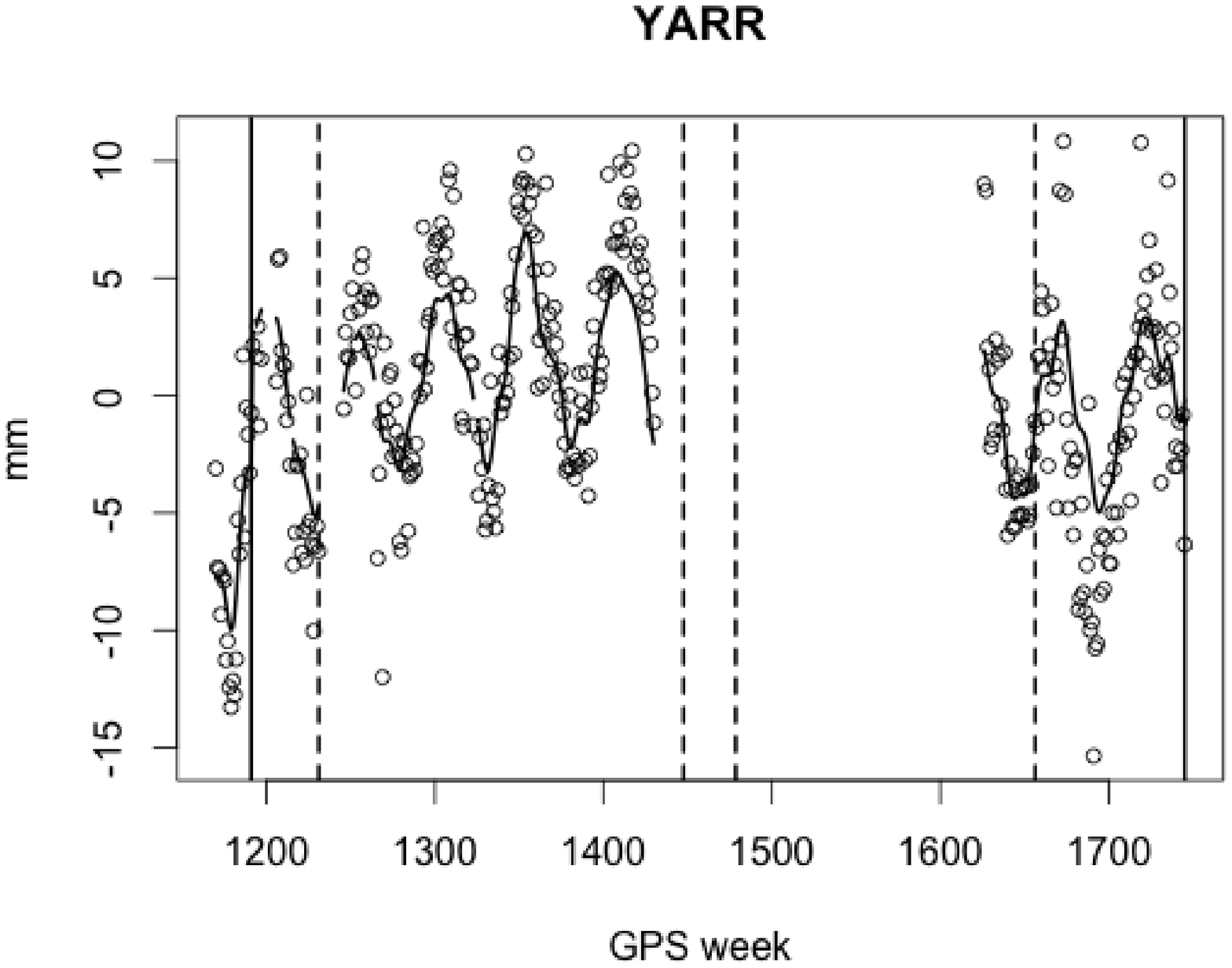}}
\end{tabular}
\caption{Results for height coordinate series of four GPS stations (YAR1, YAR2, YAR3 and YARR): obtained breakpoints in solid vertical lines; known equiment changes in dashed vertical line; estimated function $f$ in solid line.}
\label{appli}
\end{figure}

We apply our proposed procedure to these series with a dictionary with $226$ functions, which are only Fourier functions: $t\mapsto\sin\left(2\pi w_i t\right),t\mapsto\cos\left(2\pi w_i t\right)$ where $w_i=i/T$, $T=\text{max}(t)-\text{min}(t)$ and $T/i$ is larger than 8 weeks since smaller period amplitudes  are generally negligible (see in \citealp{RACD2008}). Figure \ref{appli} shows the results for the four series: the obtained breakpoints in solid vertical lines, the known equiment changes in dashed vertical line and the estimated function $f$ in solid line. \\
A total of 50 periods (62 bases) has been selected, among them the ones close to the well-known frequencies mentioned above (annual and semi-annual) and submultiples of the draconitic periods. 12 long periods - larger than 1 year - reflect well-known GPS low-frequency noise as already noticed by \citealp{Amiri2007}. \\
Heigth breakpoints are detected.  Four (GPS week $1689$ and $1707$ of the series YAR2 and $1508$ and $1559$ of the series YAR3) correspond exactly to receiver and antenna changes. The changes at time $1205$ of the series YAR2 is likely to be related to the equipment change at time $1166$. In the same series, a change at time $1628$ is detected. This change is not known from databases, however, it is also proposed by JPL. Compared to the JPL official list of changes, we found three additional changes for YAR2 at GPS week $1205$ and the two validated changes at $1689$ and $1707$. Our two other additional changes at time $900$ of the series YAR1 and at time $1191$ of the series YARR are not reported by JPL. Up to now, no explanation has been supplied for those.\\
As a conclusion, our method found the same known breakpoints as JPL official list, but includes new validated one. Moreover the $62$ bases function selected in the Lasso procedure furnish relevant geodetic information.

\section{Conclusion}\label{sec:conclu}

The proposed semi-parametric approach for the segmentation of single
or multiple series has been shown to provide a valuable and reliable
tool to assess changes and functional variations in series, as
illustrated with our GPS height series. The search for functions
that model ground motions and periodic errors here was crucial to
provide the right segmentation of the series and reliable estimates
of the breakpoint amplitudes. Conversely, because the segmentation
is simultaneous and the number and location of the breakpoints unknown, estimated functions are also more reliable. They can be
used to better interpret ground deformation observations or to
enhance the piece-wise linear coordinate model of the Terrestrial
Reference Frame (\citealp{Petit10,derm}),
widely used for geosciences and mapping applications. This would
provide a significant improvement for the users
since such coordinates are aimed to be extrapolated in the future
(up to 5 years). Because the method is totally flexible and allows
for a large number of functions to be included in the dictionary,
it could also be applied to GPS series from
active tectonic areas where the ground motion signal is more complex
and should be modeled with additional functions.

\section*{Acknowledgements}

Karine Bertin is supported by the grant ANILLO ACT--1112, CONICYT-PIA, Chile and FONDECYT project 1141258. Emilie Lebarbier is supported by the grant  CONICYT 870100003 atracci\'on de capital humano avanzado del extranjero. Cristian Meza is supported by the grant ANILLO ACT--1112, CONICYT-PIA, Chile and FONDECYT project 1141256.

\bibliography{SegLasso}
\bibliographystyle{chicago}

\end{document}